\documentclass[aps,preprint,superscriptaddress,longbibliography]{revtex4-1}
\pdfoutput=1
\usepackage{bm}
\usepackage{graphicx}
\usepackage{booktabs}
\usepackage{slashed}
\usepackage{hyperref}
\usepackage{amssymb,amsmath,amsfonts}
\usepackage{color}
\usepackage[utf8]{inputenc}
\usepackage[caption=false]{subfig}

\begin{document}

\title{A pedagagical introduction to the Lifshitz regime}

\author{Robert D. Pisarski}
\affiliation{Department of Physics, Brookhaven National Laboratory, Upton, NY 11973}
\author{Vladimir V. Skokov}
\affiliation{RIKEN/BNL Research Center, Brookhaven National Laboratory, Upton, NY 11973}
\affiliation{Department of Physics, North Carolina State University, Raleigh, NC}
\author{Alexei M. Tsvelik}
\affiliation{Condensed Matter Physics and Materials Science Division, Brookhaven National Laboratory, Upton, NY 11973}

\begin{abstract}
  We give an elementary and pedagogical review of the phase diagrams which are possible in Quantum ChromoDynamics (QCD).
  Currently, the emphasis is upon the appearance of a critical endpoint, where disordered and
  ordered phases meet.  In many models, though, a Lifshitz point also arises.
  At a Lifshitz point, three phases meet: 
  disordered, ordered, and one where spatially inhomogeneous phases arise.
  At the level of mean field theory, the appearance of a Lifshitz
  point does not dramatically affect the phase diagram.  We argue, however, that fluctuations
  about the Lifshitz point are very strong in the infrared, and significantly alter the phase diagram.
  We discuss at length the analogy to inhomogenous polymers, where the Lifshitz regime produces a bicontinuous
  microemulsion.
  We briefly mention the possible relevance to the phase diagram of QCD.
\end{abstract}

\maketitle

\section{Introduction}

Experiments at the Relativistic Heavy Ion Collider (RHIC) and the Large Hadron Collider (LHC)
have demonstrated conclusively that a new state of matter is produced in the
collisions of heavy ions at high energy.  In the central region, the system behaves like
a Quark-Gluon Plasma (QGP), at high temperature and very small baryon chemical potential.

The central region was studied originally because it is (almost) free of baryons.  This
is useful, because in themodynamic equilibrium, it is possible to compare with the results
of numerical simulations on the lattice.

It is natural to ask what happens as one goes down in energy.  In that case, even in the
central region the temperature decreases, and the baryon (or quark) chemical potential becomes
significant.  In the collisions of heavy ions it will never be possible to go to very low
temperature, but clearly the phase diagram, as a function of the temperature $T$, and the
quark chemical potential, $\mu$, is probed.

At low $\mu$ and nonzero $T$, numerical simulations on the lattice indicate that while there is no true
phase transition, there is a large increase in the pressure in a relatively narrow region
of temperature \cite{ratti_lattice_2016}.  That is, there is a crossover, which appears to
be associated with the chiral transition.

This need not remain true as the chemical potential increases.  It is plausible that at increasing
$\mu$, a line of first order transitions appears.  If so, the line of first order transitions
must end in a critical endpoint \cite{Asakawa:1989bq,Stephanov:1998dy,Stephanov:1999zu}.  This
is a true critical point, which in thermodynamic equilibrium exhibits infinite correlation lengths.

In this paper we discuss one appears to be a footnote to the phase diagram: the appearance of spatially
inhomogenous phases.  In nuclear matter this is known as pion and kaon condensates.  The appearance of
such phases is difficult to derive, even in mean field theory.  Nevertheless, although these phases
are certainly important, naively one wouldn't expect such condensates to dramatically affect the phase
diagram.

This is true at the level of mean field theory.  We show, however, that fluctuations dramatically affect
the phase diagram \cite{Pisarski:2018bct}.  In mean field theory, three phases meet at what is known as a Lifshitz point.  In
three spatial dimensions, the fluctuations at a Lifshitz point are so strong that they completely wipe
out the Lifshitz point, leaving only a Lifshitz regime.  It is possible that the critical endpoint is
completely wiped out, leaving only a line of first order transitions.  In this case, while infrared
fluctuations can be strong in the infrared, they remain finite at all points in the phase diagram.

While our arguments are qualitative, they are rather general.  We also discuss the close analogies
between the phase diagram of QCD and that in inhomogenous polymers \cite{JonesLodge,Cates12}.  There,
what we call the Lifshitz regime is known as bicontinuous microemulsions, and is of practical importance.

\section{Mean field theory: tricritical points}

We first review the standard theory of how a critical endpoint can arise.  Consider a scalar field $\phi$,
which I assume for simplicity to be single component.  It is trivial to immediate to the case where $\phi$
transforms under some global symmetry group ${\cal G}$.  We take as the Lagrangian
\begin{equation}
  {\cal L} = \frac{1}{2} \left( \partial_i \phi\right)^2 + \frac{1}{2} m^2 \phi^2 + \frac{1}{4} \lambda \phi^4
  + \frac{1}{6} \kappa \phi^6  \; .
  \label{mean_field_std}
\end{equation}
If we consider the behavior at nonzero temperature in four space time dimensions, then 
static correlations functions are determined by correlation functions 
in three spatial dimensions.  In that case, $\phi$ has dimensions of $\sqrt{}$~mass, so $\lambda$ has
dimensions of mass, while $\kappa$ is dimensionless.  Thus in the sense of the renormalization group,
$\kappa$ is a marginal operator, and should be included.

Let us begin with the case where $\lambda$ is positive.  Then we have the standard phase diagram.  The theory
is invariant under a global symmetry of $Z(2)$, $\phi \rightarrow - \phi$.  When $m^2$ is positive the
expectation value $\langle \phi \rangle = 0$, and one is in the symmetric phase.  For negative $m^2$
$\langle \phi\rangle \neq 0$, which is the broken phase.  There is a second order phase transition for
$m^2 = 0$, which is a second order phase transition.  By the renormalization group, the behavior is
controlled by the universality class of a $Z(2)$ invariant theory, such as the Ising model.  For other
models the universality class is that of the symmetry group ${\cal G}$.

It is also possible to consider negative quartic couplings, where $\lambda < 0$.  To ensure that the potential
is bounded from below, we have to assume that the hexatic coupling $\kappa$ is positive.  Then one has a first
order transition from the symmetric to the broken phase.  It is possible to determine in detail where it occurs:
there transition is from $\langle \phi \rangle = 0$ to some nonzero $\phi_0$; by the symmetry, it is to
$\pm \phi_0$.  This is determined by the potential being degenerate with $\phi = 0$, so $V(\pm \phi_0)=0$; for
$\phi_0$ to be a minimum, $\partial V(\phi)/\partial \phi = 0$ at $\phi_0$.  This is two conditions, which
can be satisfied for a given value of $\lambda < 0$ by adjusting $m^2$.  The basic point can be understood
without detailed computation: $m^2$ must be positive.  For example, at $m^2 = 0$ the potential about the origin
decreases, and so $\phi = 0$ is always above $\phi_0 \neq 0$.

\begin{figure}
\centering
\includegraphics[width=10 cm]{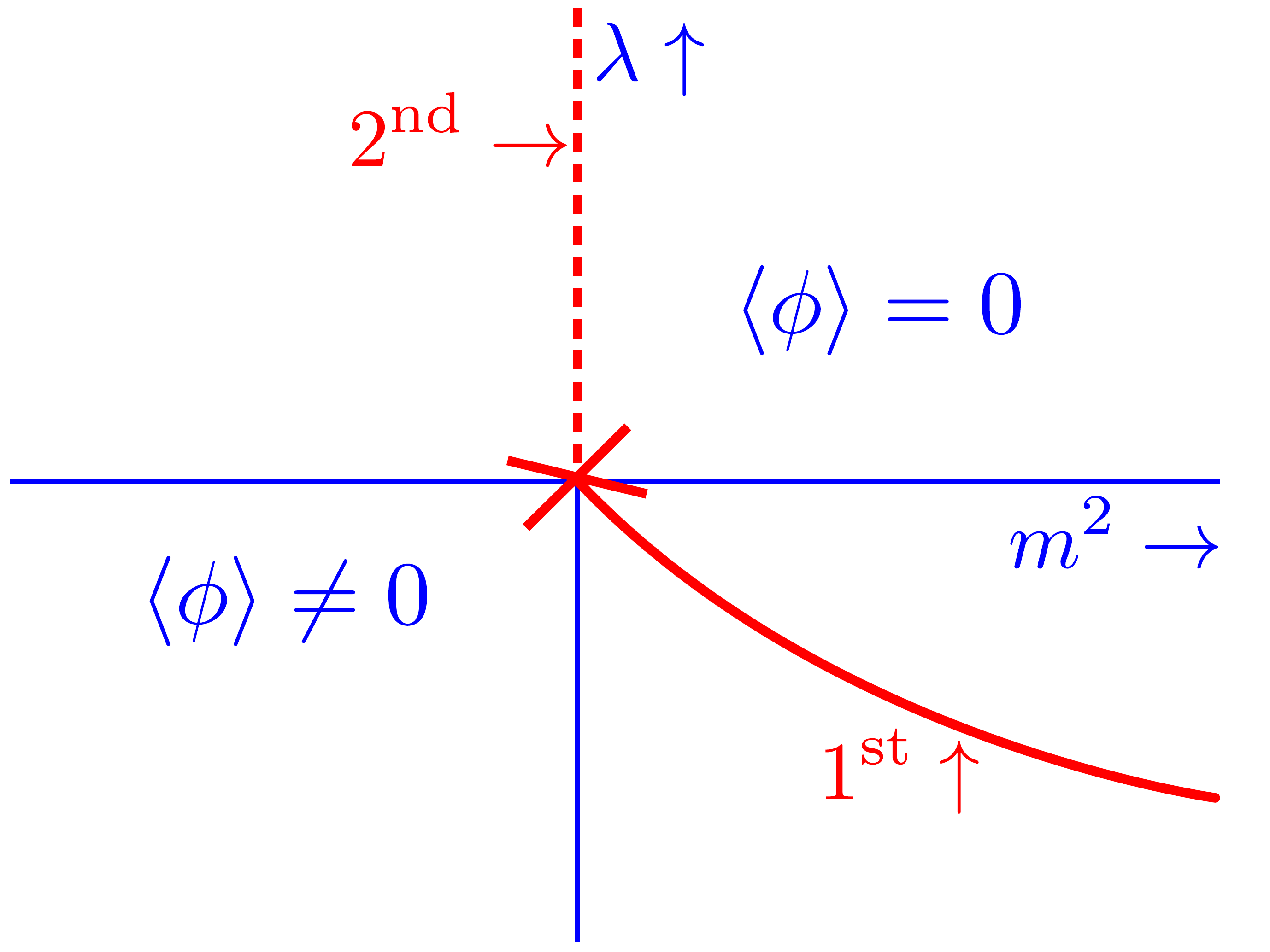}
\caption{The standard diagram in mean field theory, with a tricritical point at the origin.}
\label{mean_field_thy_std}
\end{figure}   

This gives the phase diagram of Figure \ref{mean_field_thy_std}.  There is a line of second order transitions
for $m^2 = 0$ and $\lambda > 0$, and a line of first order transitions when $m^2 > 0$ and $\lambda < 0$.
They meet at the origin, where $m^2 = \lambda = 0$.  This is a tricritical point, where both the mass
and quartic coupling vanish.  In three dimensions, the hexatic coupling runs logarithmically in the infrared,
as it is the upper critical dimension for this interaction.

If the symmetry is not exact, one adds a term which breaks the symmetry, such as $h \phi$.  If the background
field $h$ is small, the position of the transitions should be near that for $h=0$.  The most dramatic change
is that the line of second order transitions becomes a line for crossover.  That is, when $h \neq 0$ the
field always has a nonzero expectation value, $\langle \phi \rangle \neq 0$, so the theory is always in a broken phase.

For sufficiently negative values of $\lambda$, though, the first order transition must remain.  Thus the line of
first order transitions persists.  This implies that it terminates in a critical endpoint, precisely as for the
liquid-gas transition in water.  The critical field is $\phi - \langle \phi \rangle$.  If the underlying
theory has a larger symmetry of $\cal G$, the universality class of the critical endpoint remains as in the Ising
model, $Z(2)$.  This is because the field develops an expectation value along some direction, and the
critical field is still $\phi - \langle \phi \rangle$, for that particular direction.

\section{Mean field theory: Lifshitz point}

Next we consider a more general Lagrangian,
\begin{equation}
  {\cal L} = \frac{1}{2} \left( \partial_0 \phi\right)^2 + \frac{1}{2 M^2} \left( \partial_i^2 \phi\right)^2 
  + \frac{Z}{2} \left( \partial_i \phi\right)^2 + \frac{1}{2} m^2 \phi^2 + \frac{1}{4} \lambda \phi^4 \; .
  \; .
  \label{lag_lifshitz}
\end{equation}
This is an effective Lagrangian, so it is possible to have terms involving higher derivatives.
Because of causality, this is not possible for derivatives with respect to time: these can only be
to quadratic order.  For spatial
derivatives, however, it is possible to consider terms which are of higher order.  We then
include term with four spatial derivatives, $\sim (\partial_i^2 \phi)^2$; by dimensionality,
this coefficient must have dimensions of $\sim 1/M^2$, where $M$ is some mass scale which arises by
constructing the effective theory.

To ensure the theory has a stable vacuum, this coefficient must be positive.  This implies that
the usual term, with two spatial derivatives, can have a coefficient, $Z$, which is negative.
This is a circumstance which may be unfamiliar.  The theory can be either in the symmetric or
disordered phase.

\begin{figure}
\centering
\includegraphics[width=6 cm]{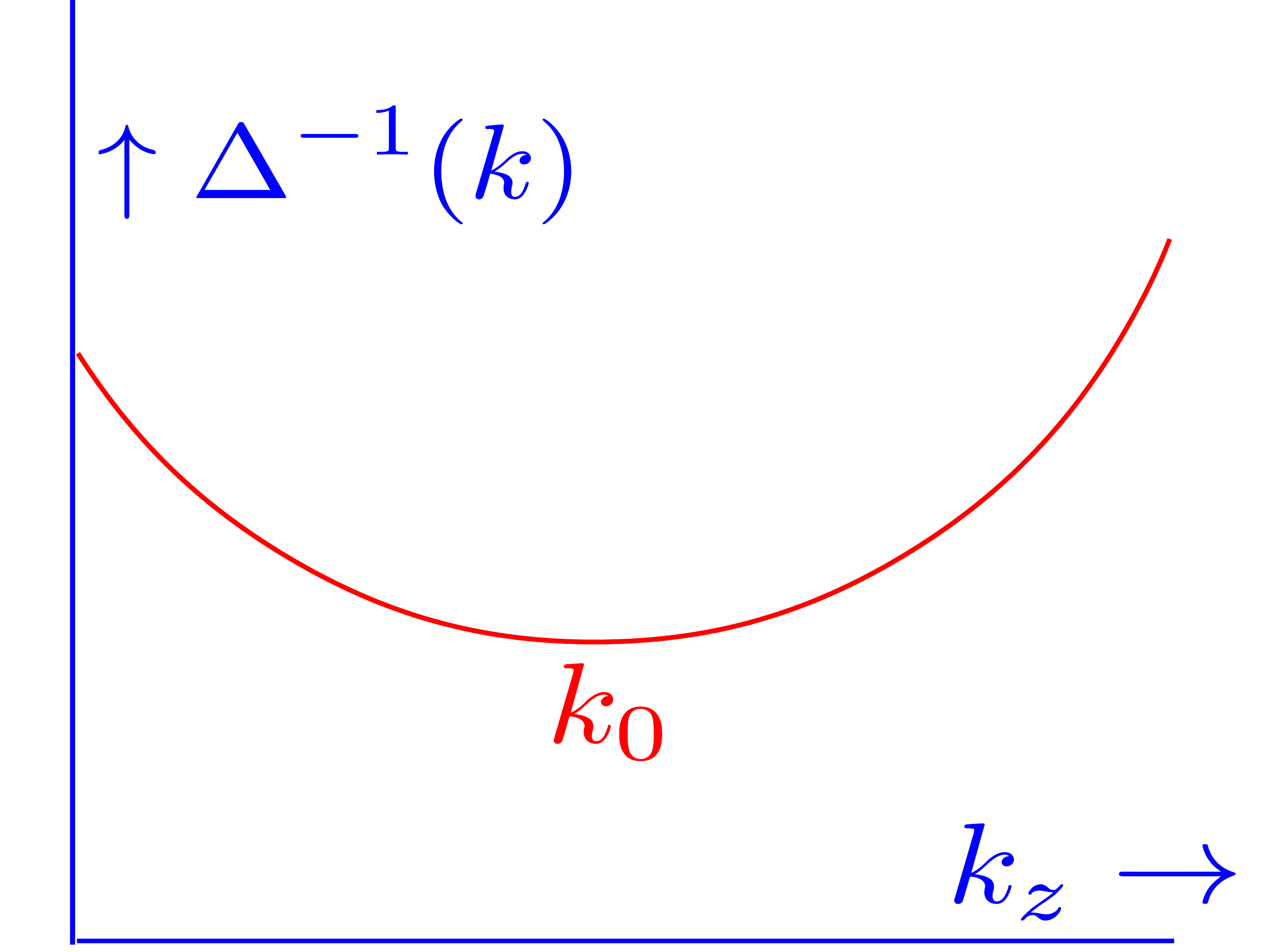}
\caption{The dispersion relation for the Lagrangian of Eq. \ref{lag_lifshitz} in the symmetric phase, where $Z < 0$ and
  $m^2 > 0$.}
\label{disp_rel}
\end{figure}   

Consider first the symmetric phase, where $m^2 > 0$.  The dispersion relation is plotted in Fig.
Figure \ref{disp_rel}.  There is no condensate, but clearly the minimum of the propagator is at
a nonzero momentum, $k_0$.  We can choose this direction to be along some direction, say $k_z$.
Expanding $\vec{k} = (k_0 + k_z,k_\perp)$, we require that the terms $\sim k_z k_0$ vanish.
The inverse propagator is then
\begin{equation}
  \frac{1}{M^2}  (\vec{k}^{\, 2})^2 + Z \, \vec{k}^2 + m^2 = 
  m_{\rm eff}^2 - 2 \, Z \, (k_z - k_0)^2 +
\frac{1}{M^2}\left( 4 k_0 k_z \vec{k}^2 + (\vec{k}^2)^2 \right)
  \;\; , \;\;
  Z < 0 \; .
  \label{disp_rel_sym}
\end{equation}
where
\begin{equation}
  k_0^2 = - \, Z \, \frac{M^2}{2} \; , \; m_{\rm eff}^2 = m^2 - \frac{Z^2}{4} \, M^2 \; .
  \label{k0_meff}
\end{equation}
The first condition is only satisfied if $Z$ is negative.   As can be seen from
the effective mass, having $Z < 0$ tends to drive the effective mass negative, but
if $m^2$ is sufficiently large, we can still remain in the symmetric phase.

It is notable that in Eq. \ref{disp_rel_sym}, the terms which are quadratic
in the transverse momenta, $k_\perp^2$, vanish identically.  This is due to the spontaneous
breaking of the rotational symmetry: the propagator has a minimum about some
nontrivial value, and we choose a direction about which to expand.  This is also why
there are terms $\sim k_z \vec{k}^2$ in the inverse propagator.  

As $Z$ becomes more negative, eventually we are driven into a phase when locally
the symmetry is broken, with $\langle \phi \rangle \neq 0$.  Since to lowest
order the kinetic term is negative, though, this is a qualitatively different state.
In detail, the nature of this state depends intricately upon the symmetry group.  We
chose to discuss the very simplest possibility, where now $\phi$ has two components.
In that case, we assume that along some arbitrary direction, which we choose to be
$\hat{z}$, that there is a spiral:
\begin{equation}
  \phi(x) = \phi_0 (\cos(k_0 z), \sin(k_0 z) ) \; .
\end{equation}
We then have two parameters to determine, $p_0$ and $\phi_0$.  The kinetic terms contribute
\begin{equation}
\frac{1}{2}  \left( \frac{k_0^4}{ M^2} + Z\,  k_0^2 \right) \phi_0^2 \; .
\end{equation}
Minimizing with respect to $k_0$ gives
\begin{equation}
  k_0^2 = - \frac{Z}{2} \, M^2 \; .
\end{equation}
This the lowest energy state with $k_0 \neq 0$ when $Z < 0$.  Using this value for
$k_0$, the value of the condensate is determined by the usual equation,
\begin{equation}
  V(\phi) = \frac{1}{2} m^2_{\rm eff} \phi^2 + \frac{\lambda}{4} (\phi^2)^2 \; .
\end{equation}
Minimizing this potential gives the usual value for the condensate,
\begin{equation}
  \phi_0^2 = - \frac{m^2_{\rm eff}}{\lambda} \; .
  \label{mean_field_cond}
\end{equation}

In this spatially homogeneous phase, while $\langle \phi \rangle \neq 0$ locally, it is not globally.
This is obvious even for one condensate oriented in a given direction, as when we integrate over $z$,
clearly $\langle \phi(z) \rangle$ will average to zero.

Further, this state is itself unstable, as we show later.  Even without computation, this can be guessed.
There is nothing special about the $\hat{z}$ direction, and fluctuations will tend to disorder the theory.
It is natural to expect that instead there is a series of patches, whose width is determined by
the underlying dynamics of the theory.  We shall discuss this later.

\begin{figure}
\centering
\includegraphics[width=10 cm]{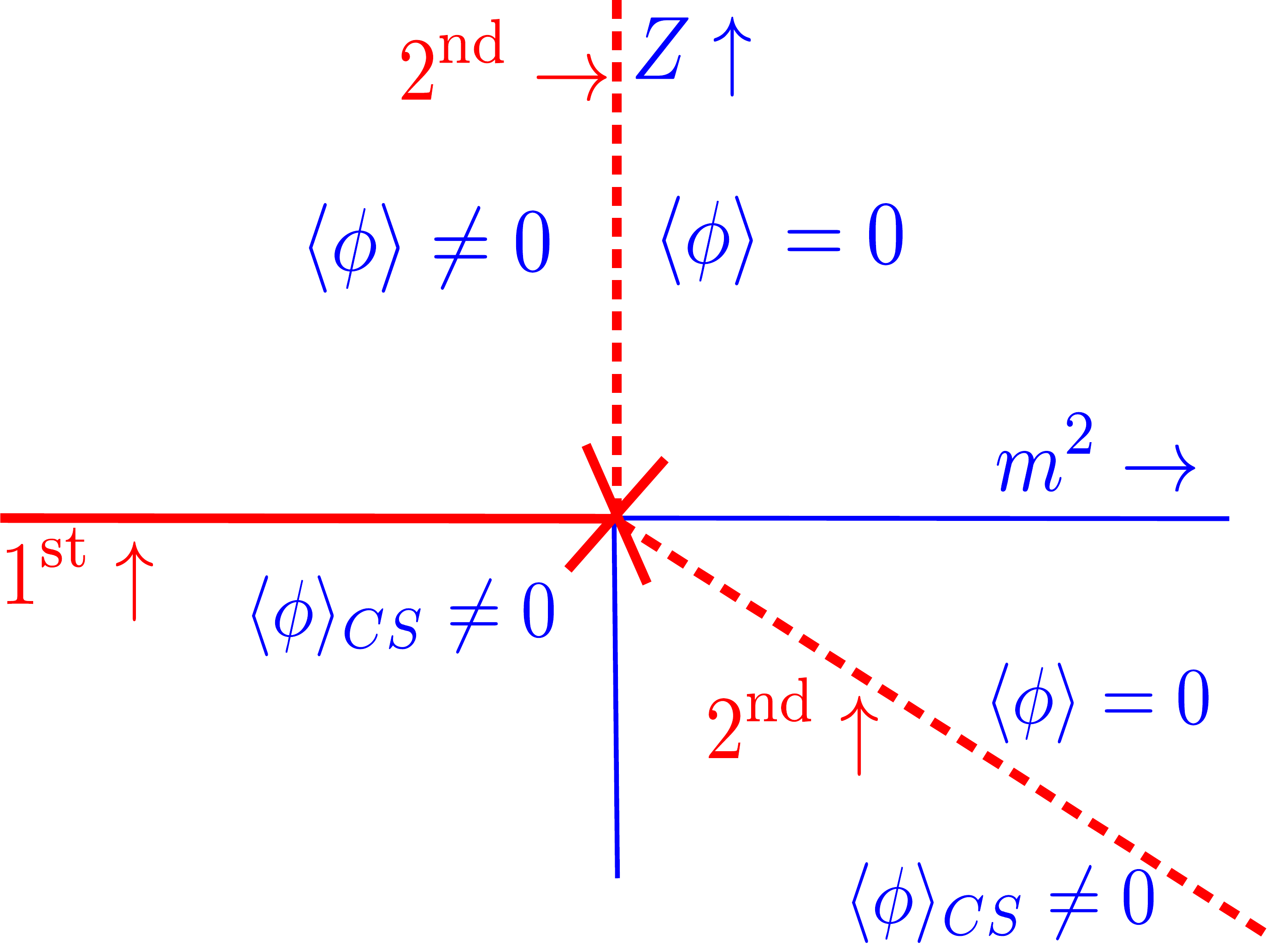}
\caption{The phase diagram for a theory with $Z < 0$ in mean field theory.  There are three phases:
  symmetry, with $\langle \phi \rangle = 0$; broken, with $\langle \phi \rangle \neq 0$, and a
  spatially inhomogenous phase, denoted CS for chiral spiral.  The three phases meet at the Lifshitz point,
  where $m^2 = Z = 0$.}
\label{lif_mean_field}
\end{figure}   

Without going into the details, we can understand the nature of the phase diagram in mean field theory,
which we illustrate in Figure \ref{lif_mean_field}.  If $Z$ is positive then we have the usual second
order transition from the symmetric to a broken phase when the mass squared vanishes.  Consider
$Z < 0$.  From the form of the effective mass in Eq. \ref{k0_meff}, it vanishes when $m^2 = Z M^2/4$.
Thus we expect a second order phase transition as we cross this line, as indicated in the figure.

When $m^2$ is negative, one goes from a broken phase, to one which is spatially inhomogenous,
as $Z$ becomes negative.  At the level of mean field theory, the free energy is given
by
\begin{equation}
  V(\phi_0) = - \frac{m_{\rm eff}^4}{4 \lambda} \; .
\end{equation}
Assume that $Z$ is linear in the temperature, $Z= a(T-T_c)$ about the critical temperature
$T_c$.  By Eq. \ref{k0_meff}, there are then terms linear in $T-T_c$ in the potential,
and so the free energy.  This is the sign of a first order transition, as then
derivative of the free energy, with respect to temperature, is discontinuous.

This is also obvious physically.  In a typical first order transition the theory jumps
from one phase to another, and the masses are discontinuous.  In this case the masses
are continuous, but the structure of the theory is completely different, as one goes
from a homogeneous ground state, to a ground state dominated by patches of spatially
inhomogeneous condensates.

\section{Anisotropic fluctuations and the phase diagram}

The phase diagram changes dramatically once fluctuations are included.  The basic
physics can be understood from the propagator in the symmetric phase, Eq. \ref{disp_rel_sym}.
Because the minimum is at nonzero momentum, the ground state spontaneously breaks Lorentz
symmetry, and the fluctuations are anisotropic.
To one loop order, there is a contribution to the mass term,
\begin{equation}
  \Delta m^2 \sim \lambda \int d^2 k_\perp dk_z \frac{1}{(k_z - k_0)^2 + m^2_{\rm eff} + \ldots}
  \sim \lambda \int^M d^2 k_\perp \int_{m_{\rm eff}} dk_z \frac{1}{(k_z - k_0)^2 }
  \sim \lambda \frac{M}{m_{\rm eff}} \; .
\end{equation}
For small effective masses, the dominant contribution is from $k_z - k_0 \sim m_{\rm eff}$, and
the anistropic propagator makes the theory effectively one dimensional.  This is
the origin of the term $\sim M/m_{\rm eff}$.  The integral over
transverse fluctuations, $k_\perp$, is cutoff by the higher order terms in the propagator,
proportional to the mass scale associated with the higher derivative terms, $\sim M$.

The effective reduction to one dimension produces the factor of $1/m_{\rm eff}$.  This implies
that while in mean field theory there is a second order transition as $m_{\rm eff} \rightarrow 0$,
this is not consistent with fluctuations.  This does not preclude a phase transition from
occurring: for a fixed, negative value of $Z$, one is clearly in a symmetric phase for large, positive
$m^2$, and in a spatially inhomogeneous phase for negative $m^2$.  Thus a phase transition must
happen, but it will do so through a first order transition, jumping from one value of $m_{\rm eff}^2$
to another.

In condensed matter physics this was first pointed out by Brazovski \cite{Brazovski:1975,Hohenberg:1995}.
There it is often referred to as a fluctuation induced first order transition, but it is rather different
from, for example, the Coleman-Weinberg phenomenon.  The latter only arises for theories with more than
one coupling constant: under renormalization group flow, one of the coupling constants flow into negative
values, thereby triggering a first order transition.  This flow is a detailed function of both the
dimensionality of space time and the symmetry group under which the fields transform.

The present case is very different: whatever the original dimensionality of space time, because of
the negative kinetic term, the infrared fluctuations are those of an essentially one dimensional theory.
Similarly, the symmetry under which the fields transform is irrelevant, all that matters is that
the fluctuations are one dimensional.  Of course the details of the transformation do depend intimately
upon these factors.  But the basic point is simply that at low momentum the theory is one dimensional,
and it is not consistent to have an interacting, massless theory in one dimension.

\begin{figure}
\centering
\includegraphics[width=10 cm]{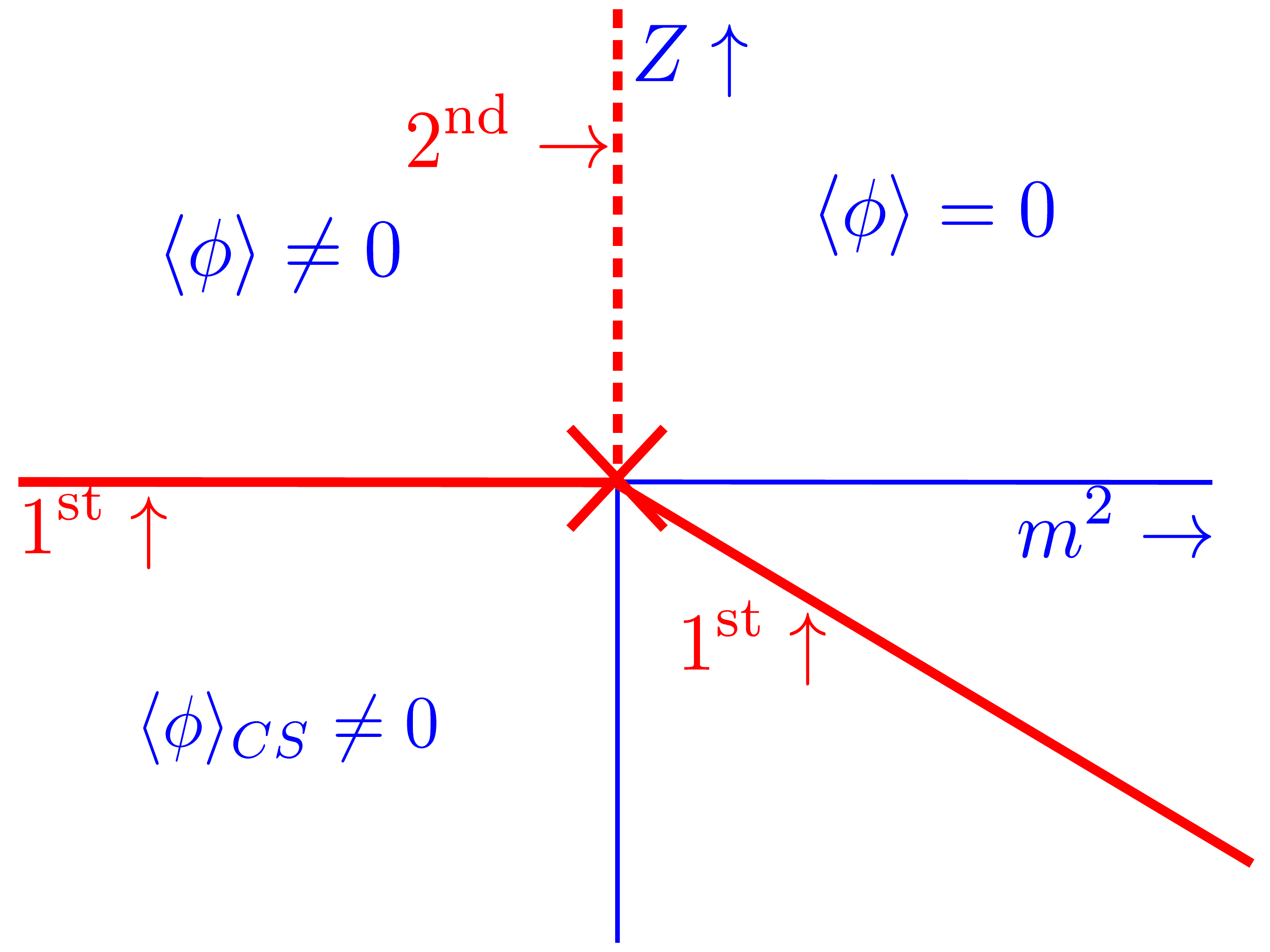}
\caption{The phase diagram for a theory with $Z < 0$, corrected for anisotropic fluctuations.
  The three phases meet at the Lifshitz point.}
\label{brazovski}
\end{figure}

This implies that the line of second order transitions, separating the symmetric and spatially
inhomogeneous phases, is in fact a line of first order transitions.  This produces the phase diagram of
Figure \ref{brazovski}.

The anisotropic form of the propagator is similar in the phase with spatial anisotropy.   Assume
that in the standard broken phase, that the theory breaks to a subgroup ${\cal H}$ of the original
group ${\cal G}$.  If $U$ are the Goldstone bosons of ${\cal G}/{\cal H}$, the effective Lagrangian
in the spatially inhomogeneous phase is
\begin{eqnarray}
  {\cal L} &=& f_\pi^2 {\rm tr} \left| ( \partial_z - i \, k_0 )U \right|^2 + \frac{c_1}{M} \; {\rm tr}
  \left( ( \partial_z - i \, k_0 ) U^\dagger  \partial_\perp^2 U + {\rm c.c.} \right) \nonumber\\
  &+& \frac{c_2}{M^2} \; {\rm tr} |\partial_\perp^2 U|^2
  + \frac{c_3}{M^2} \; {\rm tr} \left( \partial_\perp^2 U^\dagger (\partial_z - i \, k_0)^2 U + {\rm c.c.} \right)
  + \ldots \; .
\end{eqnarray}
This is the natural generalization of the propagator in the symmetric phase, Eq. \ref{disp_rel_sym}.
The coefficients $c_i$ are to be determined.  The important point is that to quadratic order in the fluctuations,
only the longitudinal momenta enter.

Consider again a tadpole diagram in this phase.  For simplicity we set $c_1 = c_3 = 0$,
\begin{equation}
  \int d^2 k_\perp \int d k_z \; \frac{1}{(k_z - k_0)^2 + (k_\perp^2)^2/M^2 } \sim
  \int d^2 k_\perp \; \frac{M^2}{k_\perp^2} \sim \log(\Lambda_{IR} ) \; ,
\end{equation}
where $\Lambda_{IR}$ is an infrared cutoff.  Because of this divergence, there is no true long range
order.  This is exactly analogous to the smectic-C phase of liquid crystals: these are systems which
are ordered in one direction, but act as a liquid in the transverse directions \cite{chaikin:2010}.

The lack of long range order is to be expected.  After all, we had assumed that the theory had
broken the three dimensional to one dimensional symmetry.  As commented, we expect that the theory
will form patches of one dimensional structure.  The interaction between the patches is controlled
by the logarithmic infrared divergences above.  Nevertheless, there can be a large separation of scales.

\section{Isotropic fluctuations}

We now consider the effect of fluctuations near the Lifshitz point.  At the Lifshitz point, the
effective theory is given by
\begin{equation}
  {\cal L} = \frac{1}{2} (\partial_0 \phi)^2 + \frac{1}{2 M^2} \left( \partial^2 \phi \right)^2 + \frac{\lambda}{4} \phi^4 \; .
\end{equation}
Although it will not enter into our considerations at nonzero temperature, which are governed by static
correlation functions, we stress that the time derivative is customary, of quadratic order.

In momentum space the propagator at the Lifshitz point is
\begin{equation}
  \Delta(k) = \frac{1}{(\vec{k}^{\,2})^2 } \; .
\end{equation}
We can now use the standard renormalization group analysis of phase transitions.   The upper critical
dimension is eight, when the renormalization of the coupling constant,
\begin{equation}
  \Delta \lambda \sim - \lambda^2 \int d^8 k \; \frac{1}{ (\vec{k}^{\,2})^2 ((\vec{p}-\vec{k})^2)^2 } \; ,
\end{equation}
develops a logarithmic divergence.  Similarly, the low critical dimension if four, when
the shift in the mass squared,
\begin{equation}
  \Delta m^2 \sim - \lambda \int d^4 k \; \frac{1}{(\vec{k}^{\, 2})^2} \; ,
\end{equation}
is logarithmically divergent.  This implies that in {\it less} than four dimensions, that there
are power like infrared divergences, and it cannot be possible to reach the Lifshitz point.

For field theories with an ordinary propagator, $\sim 1/k^2$, as is well known the upper critical
dimension is four, and the lower critical dimension, two.  The latter is familiar: it is not
possible to have interacting massless modes in two, or fewer, dimensions.

This implies that it is not possible to reach a Lifshitz point in four, or fewer, spatial dimensions.
The question is then how the phase diagram of Figure \ref{brazovski} is modified by the effects of
fluctuations.  For a theory with an exact global symmetry, there must still be the phase boundaries
indicated, with a second order phase transition between the symmetric and broken phases, and a line
of first order transitions between the spatially inhomogenous phase and the other two.  The only
difference is that it is not possible to reach the point where $Z = m^2 = 0$.

It is useful to consider the analogy to a spin system in two (or fewer) dimensions.  Begin in the
symmetric phase, and tune the mass to decrease.  Then unlike in more than two dimensions, it is
not possible to tune the mass to vanish: a nonzero mass will be generated non-perturbatively.

For a theory near the Lifshitz point, there are now two parameters to consider, $Z$ and $m^2$.
Consider moving along the line of second order phase transitions.  The mass squared must vanish
along this line.  Further, this line of second order transitions must meet the line of first order.
Consider the endpoint of this line of second order transitions.  The simplest possibility is that
$m^2$ vanishes at this endpoint.  That implies that even if in mean field theory $Z = 0$, that
$Z \neq 0$ is generated {\it non}-perturbatively.  If so, then the universality class of the
critical endpoint is the same as along the critical line.

Conversely, in mean field theory, it is only possible to go to a spatially inhomogeneous phase
when $Z$ is negative.  It is possible, however, that due to strong non-perturbative fluctuations,
that the theory develops a spatially inhomogeneous phase even if $Z$ is positive, but small.

\begin{figure}
\centering
\includegraphics[width=10 cm]{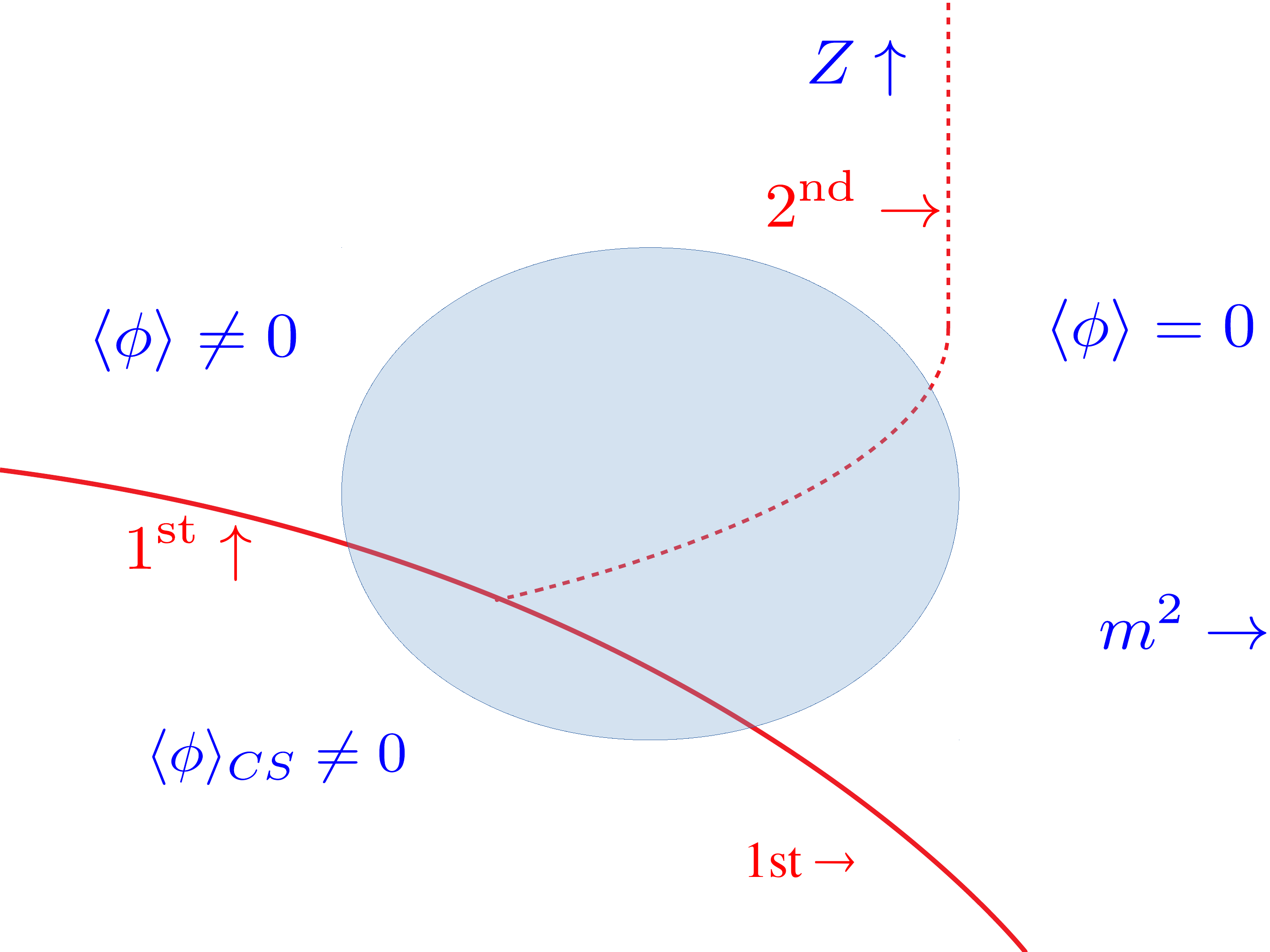}
\caption{The proposed phase diagram, including all fluctuations.  The axes $Z$ and $m$
  are not indicated, because the Lifshitz point, with $Z = m^2 = 0$, is not accessible.
  The shaded region is the Lifshitz regime, where $Z$ and/or $m^2$ are small, and infrared
  fluctuations are large.}
\label{LifshitzRegimePhaseDiagram}
\end{figure}

That is, to avoid the instability of a Lifshitz point, everywhere along the line of first order transitions
either $Z$ or $m^2$ are nonzero.  It is possible to have an isolated point where $Z = 0$, but then
$m^2$ {\it must} be nonzero.  We suggest the following: there is a region, which we term the ``Lifshitz
regime'', where $Z$ is small, and $m^2$ is nonzero.  The possible phase diagram is illustrated
in Figure \ref{LifshitzRegimePhaseDiagram}.

\section{Lifshitz regime in inhomogeneous polymers}

There is a known example of a would be Lifshitz point in inhomogeneous polymers
\cite{Fredrickson:1997,Duchs:2003,Fredrickson:2006,JonesLodge,Cates12}.
Consider first the example of a mixture of oil and water, which separate into
droplets of either oil or water.  By adding a surficant, however, the interface
tension between the phases changes, and other phases emerge.
A more controlled example is given by mixing two different types of polymers, formed of
monomers of type A and of type B, which also separate.    To this A-B diblock copolymers,
which are long sequences of type A polymer, followed by type B.  These A-B copolymers localize at the 
boundaries between phases with only A or B polymers: the part with type A sticks into the part with type A,
and similarly for type B.  The result is that adding copolymer decreases the surface tension between
the A and B phases.

The phases are then the following.  At very high temperature A and B polymers mix, which is
then a symmetric phase.  At low temperature, a mixture of A and B polymers 
separate into regions with only A or B homopolymers, which is the broken phase.
By adding the copolymer, one can obtain a lamellar phase, where A and B regions
forms stripes.  This is like a smectic liquid crystal, albeit without orientational order.

Mean field theory predicts the existence of a Lifshitz point, where these three phases meet.
In contrast, both experiment and numerical simulations with self consistent field theory
indicate that there is {\it no} Lifshitz point \cite{Fredrickson:1997,Duchs:2003,Fredrickson:2006,JonesLodge,Cates12}:
see, {\it e.g.}, Fig. 3 of Ref. \cite{JonesLodge}.

Instead, a new, intermediate region emerges near where the Lifshitz point was expected,
and is termed a bicontinuous microemulsion.  In this region
the surface tension is essentially zero.  The theory forms a a spongelike structure with large entropy,
where the polymers exhibit nearly isotropic fluctuations in composition with large amplitude.

In terms of an effective theory, the surface tension is proportional to the wave function renormalization,
$Z$.  Thus the bicontinuous microemulsion is a region where $Z$ is very small and $m^2$ is nonzero.
This is what we call the Lifshitz regime.

\section{Relation to QCD}

We conclude by briefly discussing the possible relevance to QCD.  In general, there are two possible
instabilities: either the quartic coupling constant, $\lambda$, can become negative.  This generates
the usual critical endpoint suggested previously.

It is also possible that the wave function renormalization constant for the quadratic spatial derivatives,
$Z$, becomes negative.  This generates a Lifshitz point.

At present the relationship between the two can only be studied by using effective models.  In the
simplest Nambu Jona-Lasino model, it is found that the two points coincide.  This can be understood as
following.  Starting with
\begin{equation}
  {\cal L}_{NJL} = \overline{q} \; i \!\!\not{\partial} \; q + \left( \overline{q} q \right)^2 \; .
\end{equation}
Bosonizing this through introducing $\sigma = \overline{q} q$, at one loop order we need to evaluate
\begin{equation}
  {\rm tr} \log (\not{\partial} + \sigma) \approx d_1 \left( \sigma^4+ (\partial \sigma)^2 \right) + \ldots \; ,
\end{equation}
for some constant $d_1$.  
We only indicate the first term in this expansion, as there are an infinite series of term involving higher powers
of derivatives and factors of $\sigma$.

What is found, however, is that the coefficient of the first two terms are tied together.  While surprising
at first, this can be understood through a simple scaling argument: we can rescale both length,
$\not{\partial} \rightarrow \kappa \not{\partial}$, and $\sigma \rightarrow \kappa \sigma$.  The one loop
determinant is invariant under this scaling, and so any expansion must respect it as well.

The first coefficient, $\sigma^4$, controls the location of where the quartic coupling becomes negative.
The second coefficient, $(\partial \sigma)^2$, determines when $Z$ becomes negative.  This explains why
the critical endpoint and the Lifshitz point coincide in the simplest NJL model.
See, for example, Fig. 6 of Buballa and Carignano \cite{Buballa:2014tba}.
This is also seen in solutions of Schwinger-Dyson equations
\cite{Muller:2013tya}.

This equality fails when a more complicated model is considered.  In the simplest NJL model,
This is special to models where the sigma mass is twice the constituent
quark mass, $m_\sigma/m_{\rm qk} = 2$.  Carignano, Buballa, and Schaefer \cite{Carignano:2014jla} showed that in a quark-meson
model, where one can allow $m_\sigma/m_{\rm qk} \neq 2$, that the Lifshitz and critical endpoints separate.

There are then two possibilities: in the plane of temperature $T$ and quark (or baryon) chemical potential
$\mu$, the first singularity which one meets can be either the critical endpoint, or the (would be) Lifshitz point.
Since one can only use effective models, clearly a definitive answer cannot be given.

What can be suggested is the following.  Since the simplest NJL model indicates that the critical endpoint and
(would be) Lifshitz point are near one another, it suggests the following.  The critical endpoint is dominated by
a single massless mode, which exhibit true infrared fluctuations in the infinite volume limit.  The Lifshitz
point has fluctuations which are large, but finite, in the infinite volume limit.

For the case of heavy ion collisions, which occur over a finite region of space and time, it is clearly a challenge
to distinguish between the two types of infrared fluctuations.  This is not as difficult as it may seem.  For
the critical point, there is a single massless mode, due to the $\sigma$ meson,
\begin{equation}
  \Delta_\sigma(k) = \frac{1}{k^2} \; .
\end{equation}
For the Lifshitz regime, not only the $\sigma$, but also pions and kaons exhibit a modified dispersion relation,
in which the usual quadratic propagator becomes
\begin{equation}
  \Delta_{\rm Lifshitz}(k) = \frac{1}{(k^2)^2/m^2 + m_{\rm eff}^2 } \; .
\end{equation}

\begin{figure}
\centering
\includegraphics[width=10 cm]{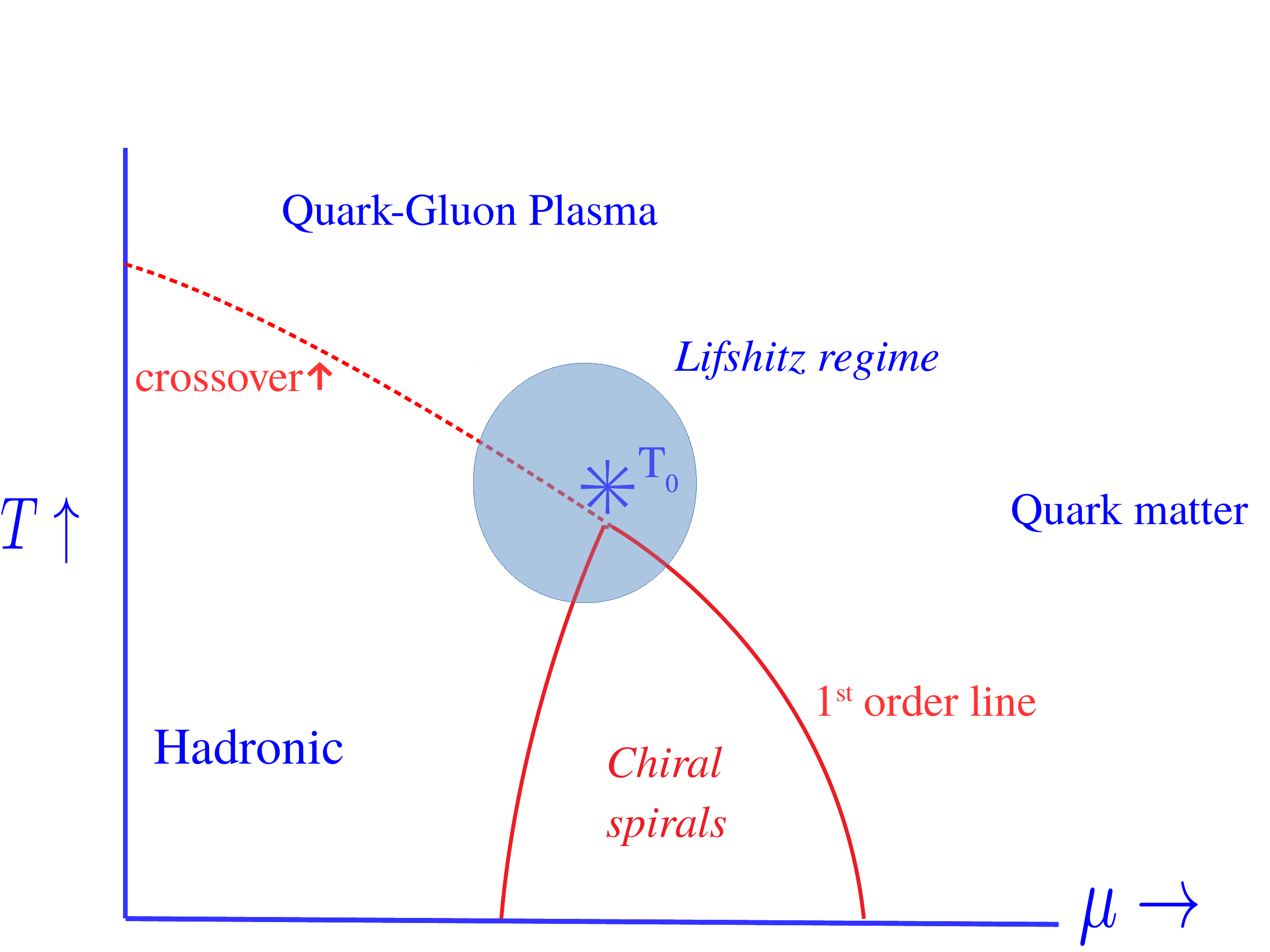}
\caption{A proposed phase diagram for QCD, in the plane to temperature $T$ and chemical potential $\mu$.
  We assume that the critial endpoint lies in the region of spatially inhomogeneous phases, or chiral spirals.
}
\label{LifshitzPhaseDiagramQCD}
\end{figure}   

A proposed phase diagram for QCD is given in Figure \ref{LifshitzPhaseDiagramQCD}.  The example
of NJL models, however, demonstrates that the existence of the Lifshitz point cannot be ignored.
Because the Lifshitz point has strong infrared fluctuations, it's presence cannot be ignored in
any analysis of a possible critical endpoint.

\maketitle

\acknowledgments
  The work of R.D.P. was supported by the U.S. Department of Energy under contract number DE-SC0012704 and by B.N.L. under the Lab Directed Research and Development program 18-036.  The of M. J., A.M.T., and R.M.K. was supported by the U.S. Department of Energy, Office of Science, Materials Sciences and Engineering Division under contract DE-SC0012704.

%\bibliography{lifshitz.bib}

\end{document}